\pdfoutput=1
\documentclass[10pt, sigconf]{acmart}

\usepackage{booktabs} 
\usepackage{graphicx}
\usepackage{balance}
\usepackage{pifont} 
\usepackage{longtable}
\usepackage{enumitem} 
\usepackage[algoruled, linesnumbered, vlined]{algorithm2e}
\usepackage{algorithmic}
\usepackage{setspace}
\usepackage{multirow}

\usepackage{sparklines}
\usepackage{xspace}

\newcommand{\ie}{i.e.,\ }
\newcommand{\eg}{e.g.,\ }

\newcommand{\tinyskip}{\vspace{3pt}}
\newcommand{\mypar}[1]{\tinyskip\noindent\textbf{#1.}\xspace}

\newcommand{\F}{\mbox{Fig.\hspace{0.25em}}}

\newenvironment{myitemize}{%
\begin{itemize}[leftmargin=1em, itemsep=.1em, parsep=.1em, topsep=.1em,
    partopsep=.1em]}
{\end{itemize}}

\newenvironment{myenumerate}{%
\begin{enumerate}[leftmargin=1em, itemsep=.1em, parsep=.1em, topsep=.1em,
    partopsep=.1em]}
{\end{enumerate}}

\newenvironment{structure*}{\color{blue}\begin{myenumerate}}{\end{myenumerate}}

\newcommand{\specialcell}[2][c]{%
  \begin{tabular}[#1]{@{}c@{}}#2\end{tabular}}

\acmConference[Arxiv]{}{March 18}{}

\begin{document}

\title{Termite: A System for Tunneling Through Heterogeneous Data}


\author{Raul Castro Fernandez, Samuel Madden}
\affiliation{CSAIL, MIT}
\email{raulcf@csail.mit.edu, madden@csail.mit.edu}

%


\begin{abstract}

Data-driven analysis is important in  virtually every modern organization. Yet,
most data is underutilized because it remains locked in silos inside of
organizations;  large organizations have thousands of databases, and billions of
files that are not integrated together in a single, queryable repository.
Despite 40+ years of continuous effort by the database community, data
integration still remains an open challenge. 

In this paper, we advocate a different approach: rather than trying to infer a
common schema, we aim to find another common representation for diverse,
heterogeneous data.  Specifically, we argue for an embedding (\ie a {\it vector
space}) in which
all entities, rows, columns, and paragraphs are represented as points. In the
embedding, the distance between points indicates their degree of relatedness.
We present Termite, a prototype we have built to \emph{learn} the best embedding
from the data.  Because the best representation is learned, this allows Termite
to avoid much of the human effort associated with traditional data integration
tasks.  On top of Termite, we have implemented a \emph{Termite-Join} operator,
which allows people to identify related concepts, even when these are stored in
databases with different schemas and in unstructured data such as text files,
webpages, etc. Finally, we show preliminary evaluation results of our prototype
via a user study, and describe a list of future directions we have identified.

\end{abstract}

%
%

%


\maketitle


\section{Introduction}
\label{sec:intro}

Data integration -- combining diverse data sets, from different organizations or
with heterogeneous schemas -- has been a long standing challenge for the
database community, which we continue to struggle with today. 
At the core of this challenge is the fact that modern relational query
processors require data to be carefully organized into a uniform schema.  In
particular, for relational operators to provide meaningful results, different
columns that reference the same concept in different data sets must use exactly
the same values and syntax.  Duplicates must be eliminated.  Values must be
normalized.  Errors must be cleaned.  Although these challenges have been a boon
to academic researchers who have published hundreds of papers on each of these
topics, they also mean that most data integration projects are hugely time
consuming and expensive, and that many data sets that should be integrated never
are due to the complexity of creating sufficiently uniform data for relational
operations to produce well-defined answers.

In this paper, we advocate an alternative approach.  Instead of insisting on
clean data and a standardized schema, we argue that we should accept that many
closely related data sets will never be fully integrated into a single
relational system. Instead, we propose Termite, a``dirt-loving'' database
system that provides as much of the power of declarative querying as possible,
but on top of these non-uniform datasets.  

The desiderata for a dirt-loving database are clear: 

\begin{myitemize} 
\item It should able to query structured but differently-schema'd tabular
data, retrieving related rows from these different tables.  
\item It should be able to relate structured to unstructured data (i.e., text
files), highlighting portions of the text files that are related to specific
records in the structured data. 
\end{myitemize}

Termite supports these goals through a novel \emph{Termite-Join} capability.
Unlike a conventional relational query processor, where most joins are based on
exact equality matches, in Termite, the join operation retrieves data that is in
\emph{close} proximity to some input query. That proximity indicates degree of
relatedness, and it is measured as the distance between vectors of an embedding
that represents all cells, rows, columns from relations as well as text from web
pages, and emails, and other files. These \emph{relational embeddings} are
similar to the word embeddings used for modelling language in text processing
\cite{we, glove}, but are specially constructed to work well for tabular
datasets that are relational in nature. 

The key advantage of the embedding representation is that because both
structured and unstructured data are represented as points (vectors),
understanding whether a tuple in a relation is related to a text file boils down
to measuring the distance between their vector representation. The key challenge
is to assign vectors to data in such a way that distance between data points in
the embedding indicates data is indeed related.  


\begin{figure}[ht!]
\centering
\includegraphics[width=0.9\columnwidth]{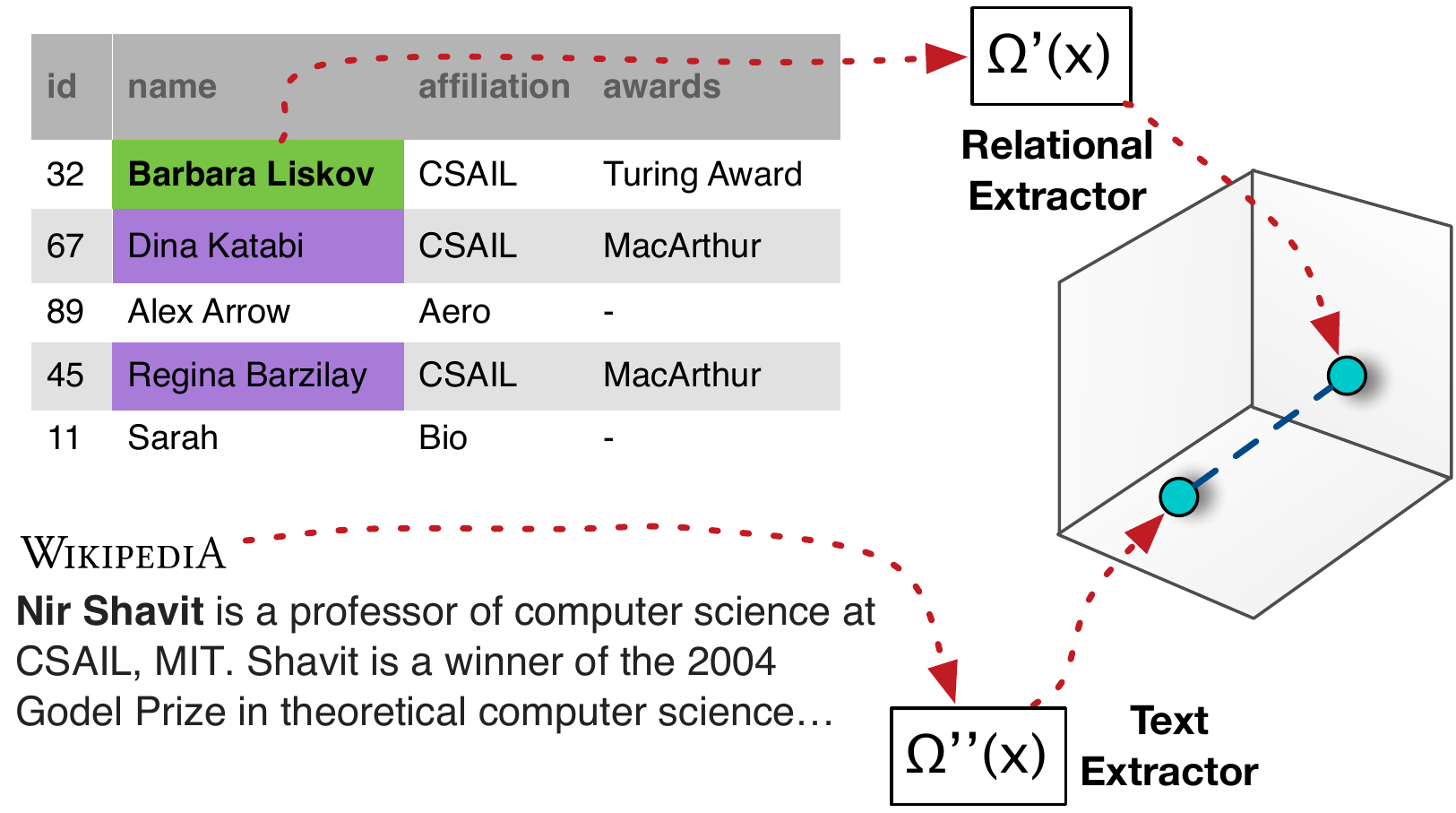}
\caption{Distance in the embedding indicates relatedness}
\label{fig:introexample}
\end{figure}

\begin{figure*}[h!]
\centering
\includegraphics[width=0.9\textwidth]{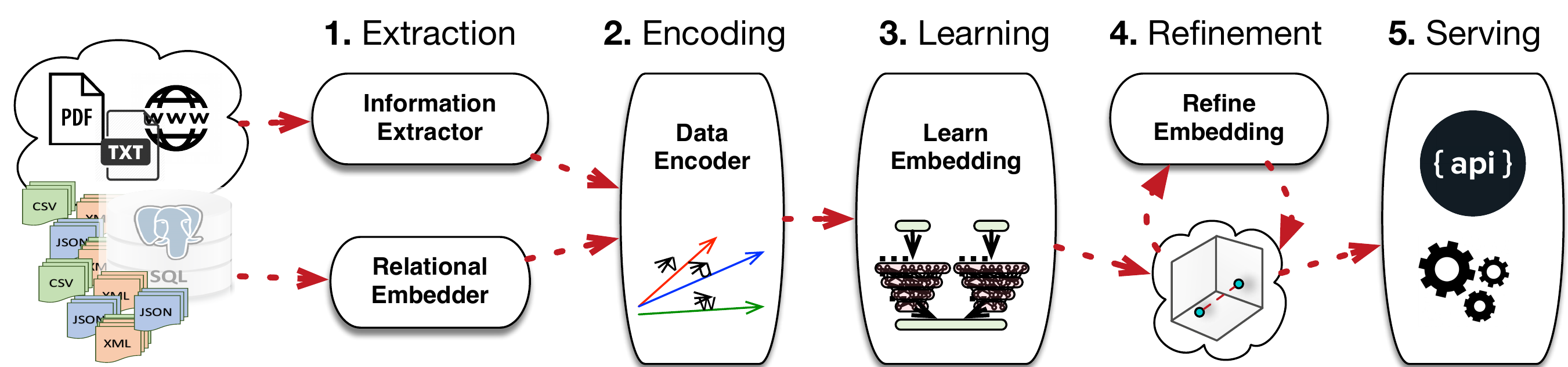}
\caption{End-to-end overview of the 5 components of the Termite system}
\label{fig:overview}
\end{figure*}

Consider the example of \F\ref{fig:introexample}, which contains a relation and
an excerpt from Wikipedia. Suppose we want to know what awards have been granted
to professors working at CSAIL. If we only looked at the table, we would miss
the Godel prize being awarded to Nir Shavit. In the embedding, the vector
representation for \emph{Godel prize} is close to the vector for \emph{Turing
award} because: i) it is an award, \ie it is \emph{granted}, or \emph{won by}
someone, and; ii) it is, in this case, awarded to someone who works at
\emph{CSAIL}. Note that a traditional TF-IDF based retrieval approach would not be able to
identify this relationship. The \emph{Termite-Join} operator relates these two
together because they \emph{share} relationships (granted/won award) and
entities (CSAIL). With Termite, we can build an embedding on which the
\emph{Termite-Join} operator works without writing manual rules. Instead,
Termite only needs pairs of elements that are related to each other; the pairs
can be generated automatically from the sentence or tuple in which the elements
appear.

\mypar{Building the embedding} A principled way of building the embedding is to
represent relational and unstructured data in some multi-dimensional
tensor---which would be very sparse---where dimensions correspond to the
different entities in the data, and then factorize the tensor to obtain a dense
embedding that would contain information about how the entities are related to
each other. This approach is so far only theoretical because we do not know how
to precisely represent all data in a tensor form, or how to factorize it in such
a way that the resulting embedding possesses the desired equivalence between
vector distance and data relatedness. Instead, we
propose to \emph{learn} the embedding directly from the data. The central theme
of this paper is \emph{Termite}, a system to train and build the embedding, and
an operator, \emph{Termite-Join} to query the embedding to relate
structured and unstructured data.

As an initial step towards Termite, we have built a proof of concept focused on
helping with discovery problems. We conducted a user study to understand
the benefits of the \emph{Termite-Join} operator to discover data across structured and
unstructured data such as MIT News, Wikipedia, personal webpages as well as
relational data from the MIT datawarehouse and DBPedia. With Termite, users
found faster more relevant content than with a baseline solution consisting of a
full-text search index carefully built. We complement our evaluation with
results on record linkage and concept expansion, two tasks closely related to
the discovery problem.

We discuss the Termite's architecture rationale in section
\ref{sec:architecture}, the current learning pipeline in \ref{sec:relemb},
followed by evaluation results (\ref{sec:evaluation}), related
work~\ref{sec:relatedwork} and a brief discussion in
\ref{sec:conclusions}.


\section{Termite's Architecture Rationale}
\label{sec:architecture}

The idea of building an embedding with Termite was inspired by the impact of
statistical language models such as word embeddings~\cite{we, glove} on the NLP,
speech recognition and information retrieval communities.  We quickly discovered
that it is not straightforward to directly use these existing techniques, mainly
because the assumption that all those models make---that words that appear often
together are related to each other---does not translate to the set-oriented
relational world of tuples, attributes, and tables, which carry much more
structured information. Furthermore, it is not clear how to merge relational data
with unstructured sources.

We have conceptualized the challenges faced by Termite into 5 loosely coupled
components (\F\ref{fig:overview}). Each stage presents a number of
research opportunities. Rather than exploring in depth each one of them, we
decided to first build an end-to-end prototype so we can learn how the different
stages are interconnected to each other. We describe these stages and our
initial implementation below: 


\mypar{Extraction} The data extraction component converts raw relations and text
into a set of bag-of-words (BoW) representation, \eg one BoW per triple
extracted from a sentence, or per cell value from a relation. To do that, it
uses different \emph{connectors}. For unstructured data, we use state-of-the-art
information extraction platforms such as \cite{openie, stanfordner, gate} to
extract entity-relationship-entity triples. For relational data, a relational
discovery tool guesses~\cite{aurum, powerpivot} each relation's key and uses it
along with the attribute values to produce triples, \eg \emph{John:value -
age:attribute - 22:value}.


\mypar{Encoding} The encoding component transforms each BoW into a vector. The
vectors must be fixed-size so they can be used as the input to the learning
component. A straightforward fixed-sized representation such as one-hot encoding
has two big drawbacks. First, its dimensionality depends on the vocabulary size,
which is large even for small datasets. Second, the vocabulary size must be
known a-priori in order to size the vectors, which is inconvenient.

Our encoding component, instead, dictionary-encodes the vocabulary terms as
integers, which are assigned incrementally as new words appear. These integers
are indexed into a fixed-sized vector of length F using a hash function 
in $1\ldots F$. We size the vector to minimize the number of collisions,
which can be achieved using the birthday paradox and the expected number of
words per BoW. Because collisions will occur anyway, we make a second attempt to
insert the integer using a different hash function. With this encoding strategy
we have seen performance improvements during learning of up to 2 orders of
magnitude compared with one-hot encoding for a vocabulary size of 1M terms.

\mypar{Learning and Refinement} These components are explained in detail in
section~\ref{sec:relemb}. Here, we only mention that given the current extractor
and encoder components, which produce triples of the form
\emph{subject-predicate-object}, the training dataset is built by generating
pairs from such triples: \emph{subject-predicate}, \emph{predicate-object} and
\emph{subject-object}. The pairs from the extracted triples are the positive
pairs. Suppose we have positive pairs that always relate a professor to a phone
number, and a phone number to an office. Even if we do not have an explicit pair
relating the professor to the the office, both entities will appear closer to
each other in the embedding: 
that's a key advantage of \emph{joining} in the embedding. 

To obtain negative samples we randomly assemble pairs that are not part of the
positive training set, similar to the approach used in~\cite{transe}.  This
makes it easy to generate negative pairs, but it introduces anomalies during the
learning process \ie unrelated points that end up close to each other not
because they are related, but because they were not explicitly provided as
negative pairs. The refinement component ameliorates some of the anomalies.

\mypar{Serving} The serving component is Termite's raison d'etre. It makes the
embedding available to answer database queries. Applications that
traditionally take most of the time from analysts who need to perform them---and
that are therefore not available to organizations without the luxury of
dedicated analysts---become straightforward to perform if an embedding is
available.

One example is data exploration, which we refer as the process of
\emph{visualizing} schemas to learn the content they represent,
\emph{summarizing} relations to get a glimpse of the information they convey,
\emph{understanding} how two relations are related to each other without going
through the process of figuring out how to join them. Each of these tasks would
take a long time to solve, but are really simple to solve in a vector space: 1)
plot vectors in a reduced dimensionality to visualize the schema; 2) find a
subset of diverse vectors to summarize a relation; 3) find vectors from the two
relations we want to join that are close to each other in the embedding.

Another example is the task of discovering how data from relations and
unstructured sources is related to each other. Useful for discovery, filling
missing values and verifying information that appears in a table among others.
This is the first application we have focused on and the reason for the
\emph{Termite-Join} operator we have implemented and focus on in the rest of
this paper.



\section{Building a Good Data Embedding}
\label{sec:relemb}

We have explained how to transform a collection of text and relations into a
collection of triples. Here, we explain how to turn the triples into a
collection of vectors, (learning component of \F\ref{fig:overview}) in
(\ref{subsec:model}). We then explain how to refine the embedding (component 4
in the figure) and briefly the \emph{Termite-Join} operator.


\subsection{Obtaining a Basic Embedding}
\label{subsec:model}

Methods such as \cite{rescal, neuraltensor, transe} consume triples from a
knowledge base to learn the latent variables that \emph{explain} observable
data, which helps among other tasks, with knowledge base completion. Our goal is
different. We want to learn a distance metric to measure the relatedness of data
coming from databases and text.

The entities we want to represent are the union of the sets of subjects, $S$,
predicates, $P$, and objects $O$ of the triples, $X = S \cup P \cup O$. We want
to find a vector representation for each entity, $f(x_i) | x_i \in X$. In
addition, given a distance function, $d()$, we want the vector representation of
two related entities, $f(x_i)$ and $f(x_j)$, to be closer to each other
according to $d()$ than to a third, unrelated vector $f(x_k)$. Finally, 
our training data consists of related and unrelated pairs.
How can we find such vector representation $f()$?

\mypar{Can't we just use Word Embeddings?} Word embeddings \cite{we, glove}
assume that words that appear often together are related to each other. Using
very large text corpora---where words are used many times in different
contexts---it is possible to learn a vector representation for each word, and
measuring the distance between word vectors, to determine whether they are
similar or not. The notion of similarity in word embeddings stems from the usage
of words in the \emph{same context}, \eg \emph{handsome} and \emph{pretty} will
be similar to each other because they are often used together in sentences.  In
our setting, 
we have the advantage of knowing
precisely which entities are related to each other (the pairs): there is no need to infer
this from their appearance together. However, we also have the disadvantage that
entities won't occur many times in many different contexts. We need to find an
alternative. 

\mypar{Vector assignment} Our proposal is to frame the assignment of vectors to
entities in $X$ as an optimization problem amenable to learning, so we can train
it efficiently by feeding the pairs we have in the training dataset. In
particular we want to train a deep network that, when given two input entities,
$x_i$ and $x_j$, assigns a vector to each of them, $f(x_i)$ and $f(x_j)$,
computes their distance and predicts whether these two entities are close to
each other. Unlike traditional machine learning models
built for generalization, \ie to predict output for unseen data, we are only
truly interested in the representation $f()$ learned by the network, so we can
use it to encode our data into the embedding. So, how do we train $f()$?


\begin{figure}[ht!]
\centering
\includegraphics[width=0.8\columnwidth]{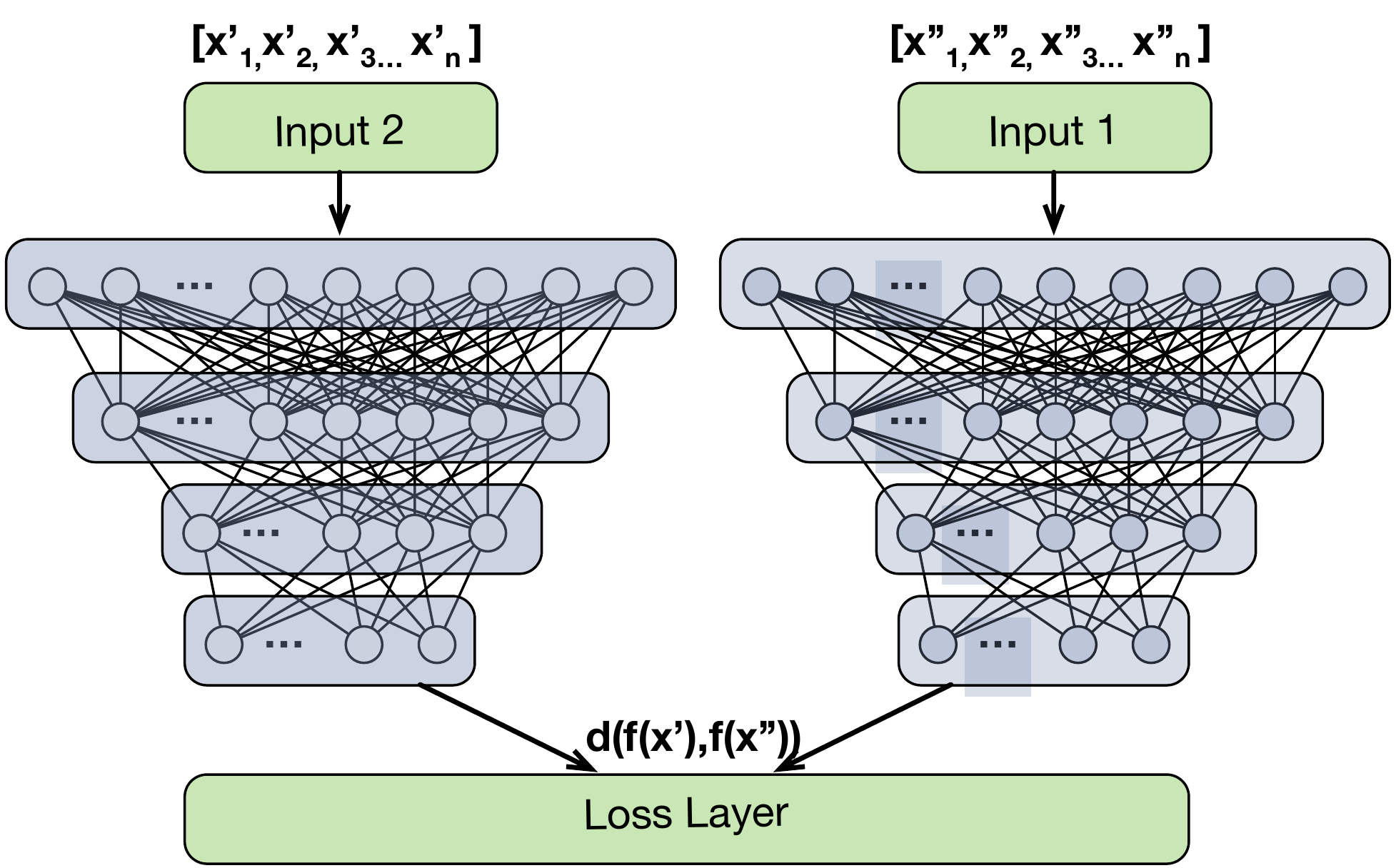}
\caption{Architecture of the siamese network}
\label{fig:siamesenetwork}
\end{figure}

\mypar{Deep metric learning} We were inspired by the siamese networks used in
\cite{siamesenetwork} for identifying images of similar faces by using a metric
learned by showing examples of similar faces, a task known as deep metric
learning. We used a network such as the one in \F\ref{fig:siamesenetwork} to
learn a metric for data. Once the network is trained and it has learned $f()$, we
apply it to each element in $X$, obtaining the embedding representation of our
data. And with that, we are back into database territory.




\subsection{Refining the Embedding}
\label{subsec:refinement}

We can use a repertoire of techniques from databases to store, index and
query the embedding efficiently. We are exploring the best ways to manage and
manipulate the embedding, but in this section we focus on how we can further
improve the embedding quality.

The learned embedding will contain anomalies: vectors that are close together
but are unrelated. This is because we only use a limited number of negative
samples during training and because of the curse of dimensionality. If we leave
the embedding untouched, we will produce wrong results when querying it. Next,
we explain the technique we have implemented in the refinement component as well
as ongoing work:



\mypar{Curse of dimensionality} Since we are working with a high-dimensional
embedding, we suffer from the curse of
dimensionality~\cite{distanceconcentration}. The worst consequence is the
phenomenon known as \emph{hubness}, which is the tendency of certain points to
be close to many other points. This 
means that certain data will be artificially related (close) to a lot of other
data, which is directly against the quality metric we desire for our embedding.

The good news is that we can largely ameliorate this problem. The main
intuition of our technique is that it is possible to compute a \emph{hubness
factor} for each entity represented in the embedding, and then remove
entities with a high hubness factor in the top-k results. In
particular, we compute how many times each point appears close to other points
in the embedding. We then take the 75 percentile of the number of appearances as
a cutoff parameter. At query time, we filter out those entities of the top-k
results with a hubness factor higher than the cutoff parameter and pad the
ranking with additional entities until we have K elements. Empirically, this
improves the quality of the returned rankings.

\mypar{Ongoing work: TDA} Can we learn more from the data once it's in a
vector format? It seems intuitively interesting to understand the shapes the
data forms in the learned embedding, and that may help us further refine the
embedding itself. Whether two or more vectors are related---whether they have a
shape---boils down to determining if they are within a specific distance,
$\delta$, of each other. Then, if they remain close as $\delta$ grows, it is
possible to determine the strength of the relationship.  Using the same
intuition, vectors that do not remain close may be categorized as noise.
Topological data analysis (TDA) is a mathematical tool that permits reasoning
about shapes in high-dimensions algebraically. A central concept in TDA is
persistent homology, which indicates which shapes remain in the high-dimensional
embedding as $\delta$ changes. A straightforward application of persistent
homology to our embedding gives us a degree of confidence for each of the
results of the top-k list, depending on how \emph{persistent} they are in the
embedding. We are currently investigating additional applications of the
technique, as well as how to best use TDA to curate the embedding.

\subsection{Termite-Join Operator}
\label{subsec:termitejoin}

Given an input entity, \emph{Termite-Join} returns the $K$ closest entities in the
embedding. All entities representations in the embedding are computed offline by
feeding the entities to the learned network and obtaining their representation.

At query time, given an input query $x_i$: 1) Obtain the embedding
representation, $f(x_i)$ from the collection. 2) Retrieve the $K$-closest
vectors to $f(x_i)$ from the collection. 3) Remove each $k_i \in K$ whose
hubness factor is beyond the cutoff parameter computed by the refinement
component. 4) Fill in the top-K list if some element has been removed in step 3.
5) Obtain the string representation of the top-k vectors and present the results
to the user.

\section{Early Experience}
\label{sec:evaluation}

We demonstrate how Termite helps users identify related data across
heterogeneous schemas and unstructured data (section~\ref{subsec:discovery}).
Then, we show additional microbenchmark results on record linkage and
concept expansion (\ref{subsec:entity}).

\mypar{Dataset and setup} We built a dataset with information of faculty at
CSAIL. The dataset contains both structured data with different schemas, \eg MIT
datawarehouse and DBPedia, as well as unstructured data, \eg Wikipedia pages,
online news articles. We then used Termite to learn the embedding, using a
laptop with 4 cores, 8GB RAM and without access to a GPU. The whole process took
around 1 hour.

\subsection{Data Discovery with Termite-Join}
\label{subsec:discovery}

We did a user study to evaluate \emph{Termite-Join}:


\mypar{Study Goals} The goals of the study were to determine: i) whether the
embedding is an appropriate abstraction to discover data across structured and
unstructured data sources; and ii) whether the semantic distance learned is
more appropriate for discovery tasks than a traditional full-text search interface
based on TF-IDF relevance.

To answer these questions, we built two different interfaces to discover data.
One of them, \textsf{Full-Text-Search (FTS)}, receives all the data from the
data extractor from \F\ref{fig:overview} and indexes it in elasticsearch
\cite{elasticsearch}. \textsf{FTS} has an API to perform keyword
queries and find the matching documents from the system. The second
interface, \textsf{Termite-Search (TMT)} is built on top of the embedding.
It is similar to the first in that
people can query with keywords, but those keywords are used as input to the
\emph{Termite-Join} operator. Our goal was to understand which interface was
better for a set of discovery tasks we describe next.

\mypar{Study Procedure} We recruited 8 users with a CS background and that are
daily users of web search engines. We asked them to solve 3 tasks. The first is
used as a training exercise, and the remaining two, (\textsf{Task 1} and
\textsf{Task 2}), are part of the experiment. We split the users in two groups
of 4 people each, and showed a different interface to each group to avoid
cross-learning effects, \ie a person learning the results with one interface and
reverse engineering the right query when using the second interface. We then
measured the coverage of the results obtained by each group and asked for their
feedback on both the questions and the interfaces they used.

We gave each user a 7 minute introduction to the corresponding interface, along
with an example walk-through to illustrate the process. An experimenter was
present at all times with the user, to clarify questions about the task goal, as
well as to suggest ways of using the API when the user had doubts. We first
explained the example task, without telling them it was an example. Users were
asked to \emph{create a list of forms of recognition (i.e., awards) that have
been given to CSAIL faculty}. We asked the users to write their results in a
text file, and explained that they could make as many search requests as they
needed. We let them use the interface for 5 minutes (we did not stop them
abruptly if they were engaged in preparing a query) and then moved on to the
next two tasks: \textsf{Task 1}: \emph{Create a list of contributions associated
with CSAIL faculty}; and \textsf{Task 2}: \emph{A list of institutions
associated with CSAIL faculty}.

\mypar{Results} The users of \textsf{TMT} were far more successful than the
users of \textsf{FTS} as the results of table~\ref{table:usresults} demonstrate.
The table shows the percentage of results found by the users when using each
interface, distinguishing between the worst result achieved by any of the 4
users using the same interface and the average one.  Remarkably, for
\textsf{Task 1}, the users that used \textsf{TMT} found a query that led to all
results. In the case of \textsf{Task 2}, only one user found all of them. We
measured the time the users took to perform each task and found that users of
\textsf{TMT} took 2 min less on average than users of \textsf{FTS}. The times
were similar, however, for \textsf{Task 2}.  

\begin{table}[]
\small
\centering
\caption{Results from User Study}
\label{table:usresults}
\begin{tabular}{|c|c|c|c|c|}
\hline
\multirow{2}{*}{}   & \multicolumn{2}{c|}{\textbf{FTS}}          &
\multicolumn{2}{c|}{\textbf{TMT}}          \\ \cline{2-5} 
                    & \specialcell{\textbf{(worst,} \\ \textbf{avg)}} & \specialcell{\textbf{Avg.} \\
\textbf{ Time}} &
\specialcell{\textbf{(worst,} \\ \textbf{avg)}} & \specialcell{\textbf{Avg.} \\ \textbf{Time}} \\ \hline
\textbf{Task 1}     & 1\%, 27\%             & 6.1m               & 100\%
& 4.2m               \\ \hline
\textbf{Task 2}     & 9\%, 18\%             & 5.1m               & 18\%, 43\%
& 5.3m               \\ \hline
\end{tabular}
\end{table}

When we asked the users for the difficulty of the different tasks they were
solving, users of \textsf{FTS} were consistent in finding \textsf{Task 1} harder
than \textsf{Task 2}, and \textsf{Task 2} harder than the \textsf{example} task.
The users of \textsf{TMT} mentioned that all three tasks were similarly simple.

\mypar{Examples} We show example results obtained by users of both interfaces in
table~\ref{table:examples} for both tasks. For \textsf{Task 1}, the users of
\textsf{TMT} found algorithmic contributions, such as \emph{LSH, Zero-knowledge
proof} when inserting software artifacts such as \emph{Vertica} or
\emph{Postgres}, see the second column of table \ref{table:examples}. Users of
\textsf{FTS} had to \emph{guess} keywords that would indicate contributions, and
this led naturally to a lower recall. Similar results were found in the case of
\textsf{Task 2}, in which users of the \textsf{FTS} had to try different
keywords such as \emph{university}, \emph{degree}, while users of \textsf{TMT}
quickly identified that using a known example, \eg \emph{Harvard} led quickly to
many other relevant results.

\begin{table*}[]
\footnotesize
\centering
\caption{Example results by users of the study}
\label{table:examples}
\begin{tabular}{|c|l|c|l|}
\hline
\multicolumn{2}{|c|}{\textbf{Task 1: Contributions of CSAIL Faculty}}
& \multicolumn{2}{c|}{\textbf{Task 2: Associated Organizations of CSAIL Faculty}}
\\ \hline
\textbf{FTS}
& \multicolumn{1}{c|}{\textbf{TMT}}
& \textbf{FTS}
& \multicolumn{1}{c|}{\textbf{TMT}}
\\ \hline
\multicolumn{1}{|l|}{\begin{tabular}[c]{@{}l@{}}arvind co founded company\\\\
Robert\_Tappan\_Morris\\ Y\_Combinator\_(company)\\\\ stonebraker focused
aurora\\\\
hari balakrishnan commercializing \\ research medusa/aurora
project\end{tabular}} & \begin{tabular}[c]{@{}l@{}}Shor's\_algorithm, karger
algorithm\\ Chord\_(peer-to-peer), haystack project\\ simultaneous
multithreading\\ Multics, VoltDB, Ingres\_(database)\\ Wait-free,
Public-key\\ Zeroknowledge\_proof\\ RSA\_(algorithm)\\
Alloy\_(specification\_language)\\ MOOC\\ \textbf{(aprox. 25 more results)}\end{tabular}
& \multicolumn{1}{l|}{\begin{tabular}[c]{@{}l@{}}morris delbarton
school\\ dewitt university michigan\\ morris harvard
university\\ demaine phd university waterloo\\ meyer
phd harvard university\end{tabular}} &
\begin{tabular}[c]{@{}l@{}}Princeton\_University\\
California\_Institute\_of\_Technology\\ University\_of\_Pennsylvania\\
University\_of\_California \_Berkeley\\ Stanford\_University\\
Colgate\_University\\ Carnegie\_Mellon\_University\\ massachusetts institute
technology\\ La\_Sapienza\_University\_of\_Rome\\ University\_of\_Michigan\\
IBM\_Almaden\_Research\_Center\\ Rice\_University, harvard university\\ \textbf{(aprox.
10 more results)}\end{tabular} \\ \hline
\end{tabular}
\end{table*}

\mypar{Conclusion} With \textsf{TMT}, users discover relevant information for
their tasks from both unstructured and structured sources more efficiently and
easier than with a full-text search index. 

\subsection{Record Linkage and Concept Expansion}
\label{subsec:entity}

Record Linkage \cite{recordlinkage} is about finding syntactically distinct
records that refer to the same real-world entity, \eg \emph{Samuel R. Madden}
and \emph{Sam Madden}. Concept Expansion \cite{conceptexpansion} is about
obtaining instances of a given concept, \eg given \emph{Harvard}, obtain
\emph{MIT}, \emph{Caltech}, \emph{Stanford}, etc. Our hypothesis is that
\emph{Termite-Join} can help with these tasks. To understand this empirically, we
retrieved ground truth for both tasks on the same CSAIL faculty dataset
mentioned above, and then implemented functions on Termite to perform each task.
We discuss the results next:

\mypar{Record Linkage} Our dataset contains 52 faculty members. The minimum
number of representations for each faculty member was 2, the average 4, and
the maximum was 7. In total there are 210 different representations for the 52
faculty. To conduct this experiment we take one representation of each member
and use it to query the embedding with the \emph{Termite-Join} operator. We then
measure how many of the found results are true alternative representations
of the original query. We could identify 77\% (163/210) different
representations used to refer to CSAIL faculty.

In particular, the Termite identifies different spellings of faculty such as
\emph{David DeWitt} and \emph{Dave DeWitt}, as well as those that include middle
names, such as \emph{David J. DeWitt}. More important, the embedding helped to
identify entities that were not syntactically similar, such as \emph{Liskov} and
\emph{Barbara Jane Huberman}. The first appears in the relational data, while
the last name only appears in an unstructured source, but since both
representations share relationships and entities they are placed closed to each
other in the embedding.

\mypar{Concept Expansion} For this experiment we compiled ground truth following
a procedure similar to the one described in \cite{conceptexpansion}. We
found instances of the same concept for 10 different concepts, which are
shown in the first column of table~\ref{table:concept-table}. The concepts had a
number of instances that ranged from 10 to 80.  Unlike the original definition
of the concept expansion problem \cite{conceptexpansion}, which takes
both a concept and an example instance as input, we only provide an example
instance to the \emph{Termite-Join} operator. We then run queries retrieving lists
of different sizes, 2 and 4 times the size of the original (referred to as 2x
Top and 4x Top in the table), and reporting the total percentage of values in
the ranking result that were correct. The first query returns a list of size
equal to the number of instances for the concept. The second and third query
subsequently double the previous size.

\begin{table}[]
\small
\centering
\caption{Results for concept expansion}
\label{table:concept-table}
\begin{tabular}{|c|c|c|c|c|}
\hline
\textbf{Concept}                   & \textbf{\specialcell{\# \\ Instances}} &
\textbf{\begin{tabular}[c]{@{}c@{}}Found\\ Top\end{tabular}} &
\textbf{\begin{tabular}[c]{@{}c@{}}Found\\ 2x Top\end{tabular}} &
\textbf{\begin{tabular}[c]{@{}c@{}}Found\\ 4x Top\end{tabular}} \\ \hline
\textit{\textbf{\specialcell{known \\ for}}}         & 77                    &
\begin{tabular}[c]{@{}c@{}}27\\ 35\%\end{tabular}            &
\begin{tabular}[c]{@{}c@{}}46\\ 59\%\end{tabular}               &
\begin{tabular}[c]{@{}c@{}}65\\ \textbf{84}\%\end{tabular}               \\ \hline
\textit{\textbf{\specialcell{faculty \\ name}}}      & 52                    &
\begin{tabular}[c]{@{}c@{}}30\\ 57\%\end{tabular}            &
\begin{tabular}[c]{@{}c@{}}46\\ 88\%\end{tabular}               &
\begin{tabular}[c]{@{}c@{}}51\\ \textbf{98}\%\end{tabular}              \\ \hline
\textit{\textbf{\specialcell{Institutions}}}     & 44                    &
\begin{tabular}[c]{@{}c@{}}26\\ 59\%\end{tabular}            &
\begin{tabular}[c]{@{}c@{}}33\\ 75\%\end{tabular}               &
\begin{tabular}[c]{@{}c@{}}37\\ \textbf{84}\%\end{tabular}               \\ \hline
\textit{\textbf{\specialcell{birth \\ place}}}       & 35                    &
\begin{tabular}[c]{@{}c@{}}14\\ 40\%\end{tabular}            &
\begin{tabular}[c]{@{}c@{}}18\\ 51\%\end{tabular}               &
\begin{tabular}[c]{@{}c@{}}24\\ 68\%\end{tabular}               \\ \hline
\textit{\textbf{award}}            & 33                    &
\begin{tabular}[c]{@{}c@{}}19\\ 57\%\end{tabular}            &
\begin{tabular}[c]{@{}c@{}}24\\ 72\%\end{tabular}               &
\begin{tabular}[c]{@{}c@{}}28\\ \textbf{84}\%\end{tabular}               \\ \hline
\textit{\textbf{\specialcell{academic \\ children}}} & 54                    &
\begin{tabular}[c]{@{}c@{}}41\\ 75\%\end{tabular}            &
\begin{tabular}[c]{@{}c@{}}53\\ \textbf{98}\%\end{tabular}      &
---      \\ \hline
\textit{\textbf{field}}            & 21                    &
\begin{tabular}[c]{@{}c@{}}13\\ 61\%\end{tabular}            &
\begin{tabular}[c]{@{}c@{}}14\\ 66\%\end{tabular}               &
\begin{tabular}[c]{@{}c@{}}16\\ 76\%\end{tabular}               \\ \hline
\textit{\textbf{nationality}}      & 10                    &
\begin{tabular}[c]{@{}c@{}}10\\ \textbf{100}\%\end{tabular}           &
--- &
---                                                                  \\ \hline
\textit{\textbf{\specialcell{doctoral \\ advisor}}}  & 31                    &
\begin{tabular}[c]{@{}c@{}}12\\ 38\%\end{tabular}            &
\begin{tabular}[c]{@{}c@{}}16\\ 51\%\end{tabular}               &
\begin{tabular}[c]{@{}c@{}}23\\ 74\%\end{tabular}               \\ \hline
\textit{\textbf{\specialcell{thesis title}}}      & 19                    &
\begin{tabular}[c]{@{}c@{}}8\\ 42\%\end{tabular}             &
\begin{tabular}[c]{@{}c@{}}13\\ 68\%\end{tabular}               &
\begin{tabular}[c]{@{}c@{}}16\\ \textbf{84}\%\end{tabular}               \\ \hline
\end{tabular}
\end{table}

\mypar{Conclusions} The best news is that we obtained these results by only
pointing out Termite to a data repository. No manual domain-specific engineering
was necessary.

\section{Related Work}
\label{sec:relatedwork}


\mypar{Automatic Knowledge Base Completion} RESCAL~\cite{rescal} models triples
from a knowledge base via the pairwise interactions of latent features.
Similarly, Structured Embeddings and subsequent work~\cite{structuredembeddings,
transe, neuraltensor} learns embeddings for each relation from the triples.
These approaches focus on learning the latent variables that describe the
triples, to later fill in values of an incomplete knowledge base. In contrast,
Termite learns a metric we use to relate structured and unstructured data based
on their distance in the embedding.


\mypar{Universal schema} \cite{universalschema, rowless} decomposes a matrix in
which rows represent entity pairs and columns represent relations between
entities. Similar to our embedding, they show how to jointly embed text and
knowledge bases. They are not focused on data discovery, but rather information
extraction applications. 



\mypar{Word Embeddings and Relational Data} In \cite{wedb}, the authors propose
a method to learn a vector representation of data items from relational data
based on word embeddings \cite{we}, and then use those vectors to augment
traditional SQL queries with so-called \emph{cognitive} capabilities such as
finding elements within a column or row that are similar to an input data item.
In contrast, our embedding's goal is to find a relatedness metric
for discovery applications beyond only relational data. Our embedding could be
applied to extend SQL queries as well, which is interesting future work.

\mypar{Other related techniques} There are a myriad of applications in NLP which
share some characteristics with out goal of learning a good embedding of both
structured and unstructured data. We can benefit from: i) sequence
learning~\cite{lstm, languagetranslationlstm}, ii) alternative deep metric
methods~\cite{triplet, multipair}; iii) alternative embedding methods such as
holographic~\cite{holographic} and hyperbolic embeddings~\cite{hyperbolic}.
A recent paper~\cite{deeper} explores deep learning for entity resolution, a
common data integration task.
Termite's focus is on leeting users operate on 
structured and unstructured data without a high upfront cost.

\section{Research Agenda}
\label{sec:conclusions}

In this paper we have shown how, by operating on a vector space, we can
understand relationships between structured and unstructured data sources. We
have evaluated Termite, our proof of concept, with a few common applications in
data integration as well as with a user study. The applications of embedding
data into a common vector space, as well as the challenges of doing so, however,
span beyond what we presented here. We use the remainder of this section to
present challenges and additional applications:

\begin{myitemize}
\item\textbf{Learning a relational embedding.} 

The siamese network presented in this paper permits merging relations and text
easily, but it is slow to train. We have performed
preliminary results with neural network architectures designed for modeling
language, such as Skipgram and continuous bag of words (CBOW), which are simpler
and therefore faster to train.  Adapting these models to work with structured
data is not trivial because they have been designed to use the sequence nature
of text and not the row- and column-wise relationships represented in tables.
One important challenge, then, is to understand the different learning solutions
and identify tradeoffs between different models. 
\item\textbf{Cleaning embeddings. } All of the learning methods we 
consider lead to high dimensional vector spaces. Operating on these spaces is
challenging for several reasons. First, we suffer from the hubness
problem~\cite{hubness}, where
spurious vectors appear to be close to most other vectors. This behavior
interferes with out goal of relying on distance between vectors to understand
properties of the underlying data. A challenge in this setting is
to detect and clean hubs from the embedding efficiently. For that, we are exploring
techniques from shape analysis as well as topological data
processing, which we think may help us identify and correct errors in data.
\item\textbf{Reducing storage overhead} Because every token is mapped to a high
dimensional vector, we suffer a severe storage overhead (sometimes up to 20x)
because we need to store 
many floating points in addition to the original token. 
We cannot learn a low dimensional embedding directly because we need 
high capacity networks to learn good embeddings. However, once we possess the
high dimensional embedding, it is possible to apply dimensionality reduction
techniques to reduce the storage overhead. The challenge at this point is to
apply the dimensionality reduction techniques on millions of vectors, a process
that becomes too slow. To tackle this problem, we are exploring techniques to
reduce the dimensionality of a sample of the data only, and then derive vectors
in a low dimensional space by taking into consideration the relative distance of
points in the original embedding.
\item\textbf{Identifying intrinsic evaluation criteria.} In this paper we have
measured the quality of the learned embedding by using downstream tasks that are
well defined, and for which we can compare metrics easily. An open question is
whether we can identify features of vector embedding we can use to perform
intrinsic evaluations, which would help with speeding up the rate of innovation.
\end{myitemize}

Despite the challenges, we are motivated to pursue this line of work,
not only because of the promise of merging structured and unstructured data, but
also because the promise of many other applications which we briefly discuss
here:

\begin{myitemize}
\item\textbf{Database exploration. } Many techniques exist to explore databases,
from clustering of tables, to summarizing relations, to visualizing what tables
look more like each other, and many more. Each of these applications requires the
design an implementation of a system. We are exploring how we can translate all
the previous applications into mere vector operations, and therefore simplify
the database exploration process end to end.
\item\textbf{Statistics retrieval. } Databases, and in particular, query
optimizers, rely on data statistics to perform well their job. A relational
embedding captures data properties that can be beneficial to databases.
\item\textbf{Performance benchmarking. } When customers of commercial databases
have a performance problem and the vendors are not allowed to see the data,
troubleshooting the underlying problem becomes difficult. To tackle this problem
we can train a generative model from the embedding, and then use the generative
model to create databases of different sizes for benchmarking. Furthermore, we
can prevent certain security risks by obfuscating the tokens in the dictionary
that maps them to vectors. 
\end{myitemize}

We believe the vision exposed in this paper along with the encouraging results
justify investing more time in understanding how a vector representation of data
may benefit difficult data management problems.

\bibliographystyle{ACM-Reference-Format}
\balance
\bibliography{ms}



\end{document}